\documentclass[
aps,%
12pt,%
final,%
notitlepage,%
oneside,%
onecolumn,%
nobibnotes,%
nofootinbib,%
superscriptaddress,%
noshowpacs,%
centertags]%
{revtex4}

\usepackage{graphicx}
\usepackage{epstopdf} 
\usepackage{amssymb}
\usepackage{amsmath}
\begin{document}
%
%
%
\title{Spatial Structure of the  $^{12}$C Nucleus in a  3$\alpha$ Model with Deep Potentials Containing Forbidden States}
%
%

\author {E. M. Tursunov}
\email{tursune@inp.uz}
 \affiliation {Institute of Nuclear Physics,
Academy of Sciences, 100214, Ulugbek, Tashkent, Uzbekistan}
\affiliation {National University of Uzbekistan, 100174, Tashkent,
Uzbekistan}
\author{M. Z. Saidov}
\email{matlubsaidov7@gmail.com} \affiliation {Institute of Nuclear Physics,
Academy of Sciences, 100214, Ulugbek, Tashkent, Uzbekistan}
\author{M. M. Begijonov}
\email{marufjon19980310@gmail.com} \affiliation {Institute of Nuclear Physics,
Academy of Sciences, 100214, Ulugbek, Tashkent, Uzbekistan}

\begin{abstract}
The spatial structure of the lowest 0$_1^+$, 0$_2^+$, 2$_1^+$  and 2$_2^+$ states
of the $^{12}$C nucleus is studied within the 3$\alpha$ model with the Buck, Friedrich and Wheatley $\alpha \alpha$ potential with Pauli forbidden states in the $S$ and $D$ waves. The Pauli forbidden states in the three-body system are treated by the exact orthogonalization method. The largest contributions to the ground  and excited  2$_1^+$  bound states energies come from the partial waves $(\lambda, \ell)=(2,2)$ and $(\lambda, \ell)=(4,4)$. As was found earlier, these bound states are created by the critical eigen states of the three-body Pauli projector in the 0$^+$ and 2$^+$ functional spaces, respectively. These special eigen states of the Pauli projector are responsible for the quantum phase transitions from a weakly bound "gas-like" phase to  a deep "quantum liquid" phase. In contrast to the bound states, for the Hoyle  resonance 0$_2^+$ and its analog state 2$_2^+$, dominant contributions come from the $(\lambda, \ell)=(0,0)$ and $(\lambda, \ell)=(2,2)$  configurations, respectively. 
The estimated probability density functions for the $^{12}$C(0$_1^+$) ground and 2$_1^+$ excited bound states show mostly a triangular structure, where the $\alpha$ particles move at a distance  of about 2.5 fm from each other. However, the spatial structure of the Hoyle resonance and its analog state have a strongly different structure, like $^8$Be + $\alpha$.  In the Hoyle state the last $\alpha$ particle moves far from the doublet at the distance between $R=3.0$ fm and $R=5.0$ fm. In the Hoyle analog 2$_2^+$ state the two alpha particles move at a distance of about 15 fm, but the last $\alpha$ particle can move  far from the doublet at the distanse up to $R=30.0$ fm.
\end{abstract}

 \maketitle

\section{Introduction}
\par  A discovery of  quantum phase
 transition (QPT)  in the $\alpha$-like nuclei \cite{serdar16}, including $^{12}$C ground state,
 is one of the main phenomena in nuclear structure studies of last years.
 The effect was found within the framework of {\it ab initio}
 method based on the chiral effective field theory
potentials. The importance of the finding is connected with a role
of the carbon element in the Universe, thus allowing to conclude
about "the life near a quantum phase transition" \cite{serdar16}.
The {\it ab initio} method of \cite{serdar16} is of course a
powerful tool for studying hyperfine effects from first principles
of the quantum chromodynamics (QCD). In this sense the method has
more predictive power than other approaches (see recent review
\cite{rev} and references therein).

The discovery of the QPT  in $^{12}$C inspired new research interest
on the structure of this important quantum object. Very important
question is, whether it is possible to observe an effect of the QPT
within the framework of a 3$\alpha$-cluster model for practical
applications. Another interesting property of this nucleus is its
special structure, associated with the Bose--Einstein condensation
\cite{review12C}. On the other hand  a realistic modeling of Pauli
forces in the 3$\alpha$ system is a very difficult problem. A complicated nonlocal $\alpha \alpha$ potential
derived from the resonating group model calculations was able to
reproduce the energies of the ground state and the Hoyle ($0^+_2)$
resonance \cite{suz08}. But a repulsive local $\alpha \alpha$-potentials, both l-dependent
and l-independent, strongly underestimate the bound states of the
$^{12}$C nucleus \cite{tur03}. The alternative
local deep $\alpha\alpha$- interaction potential of
Buck--Friedrich--Wheatley (BFW) \cite{BFW} yields much more realistic
description of the nuclear structure, since it allows to treat the Pauli forbidden
states (FS)  exactly. However, application of the  method of orthogonalizing pseudopotentials (OPP) \cite{kuk78}
for the 3$\alpha$ quantum system met a serious problems, although it works very well for the structure of nuclei like $^6$He and $^6$Li, which contain a single  $\alpha$ cluster  in the $\alpha+N+N$ three-body model \cite{tur06,tur06a,tur16,bt18,tur18} .

Indeed, detailed studies within the OPP method  have demonstrated \cite{tur01,fuji06} that the energy spectrum
of the ground $0_1^+$ and first excited $2^+_1$ states is highly
sensitive to the description of the $\alpha \alpha$-Pauli forbidden
states. In these studies a convergence of the  energies of the ground and excited states
in respect to the projecting constant $\lambda$
was not clear. When passing values $\lambda=10^3$--10$^4$ MeV the
energies of the ground $0_1^+$ and first excited $2^+_1$ states
increase sharply, which was not usual.  Only the use of the direct orthogonalization method \cite{fuji06} for the elimination of
the 3$\alpha$ Pauli forbidden states allowed to clarify the situation. It was found that the convergence problem within the OPP method  is due to the so-called almost forbidden states (AFS) which are the special eigen states of the complete 3$\alpha$ Pauli projector with eigen values, close to zero.

Finally, recently in \cite{tur21a} it was argued that above AFS are nothing but the critical eigen states (CES) of the three-body Pauli projector which are responsible for the quantum phase transition in the ground and first excited 2$_1^+$ bound states of the $^{12}$C nucleus found in \cite{serdar16}.  So, when passing the quantum critical eigen state, the energy of the lowest state of the carbon nucleus changes sharply from -0.627   MeV (weakly bound phase) to the deep phase with the energy value of -19.897 MeV.  The same behavior was found for the excited  $^{12}$C (2$_1^+$) bound state where the weak phase with the energy of 1.873 MeV changes to the deep phase with the energy value of -16.572 MeV.  Another interesting result is that from the left side of the quantum critical point the lowest states present the astrophysical significant Hoyle state 0$_2^+$ and its excited analog state 2$_2^+$. When passing the above mentioned CES these Hoyle states energies almost do not change, but they become the excited states. In other words the lowest   0$_1^+$ and  2$_1^+$ states in the deep phases are created by the critical eigen states of the complete three-alpha Pauli projector.

The aim of the present work  is to study  the spatial structure of the ground and lowest excited states of the
 $^{12}$C nucleus in the 3$\alpha$ model. In \cite{sam22} the spatial structure was studied within the local $\alpha \alpha$ potentials with a strong repulsive core, while adding a very large three-body attractive forces.
Differently, we deal with a more realistic deep BFW $\alpha \alpha$ potential with forbidden staters in the $S$ and $D$ waves.
A variational method on symmetrized Gaussian basis is employed. For the elimination of the
3$\alpha$ Pauli forbidden states we use the same direct
orthogonalization method from \cite{fuji06,tur21a}. We will examine a
similarity of the  $0^+$ and $2^+$ spatial structure including the Hoyle
band.

The theoretical model is described in Section 2. Sections 3
contains the numerical results for the $^{12}$C(0$^+$) and
$^{12}$C(2$^+$) states. Conclusions are given in the last section.

\section{Theoretical model}

The direct orthogonalization method \cite{fuji06,tur21a} is based on the
separation of the complete Hilbert functional space into two parts.
The first allowed subspace $L_Q$ is defined
by the kernel of the 3$\alpha$ projector. The rest
subspace $L_P$ contains 3$\alpha$ states forbidden by the Pauli
principle. The complete Hilbert functional
space of 3$\alpha$ states is separated into the $L_Q$ (allowed) and
$L_P$(forbidden) subspaces, then the three-body
Schr\"{o}dinger equation is solved in $L_Q$.

The $\alpha\alpha$- interaction potential of Buck--Friedrich--Wheatley
\cite{BFW} of the Gaussian form reads %
\begin{equation}
V(r)=V_0 \exp(-\eta r^2)+4e^2erf(br)/r,
\end{equation}
with parameters $V_0$ = -122.6225 MeV, $\eta=0.22$ fm$^{-2}$ for the
nuclear part and  $b$ = 0.75 fm$^{-1}$ for the Coulomb part. This
choice of the potential parameters yields a very good description of
the experimental phase shifts $\delta_L(E)$ for the $\alpha \alpha$-
elastic scattering in the partial waves $L=0,2,4$ within the energy
range up to 40 MeV and the energy positions and widths of the
$^8$Be resonances.

As in \cite{tur21a,fuji06} we use a value $\hbar^2/m_{\alpha}=10.4465$ MeV fm$^2$. This potential
contains two Pauli forbidden states in the $S$ wave with energies
$E_1=-72.6257$ MeV and $E_2=-25.6186$ MeV, and a single forbidden
state in the $D$ wave with $E_3=-22.0005$ MeV. A realistic
description of the system requires to eliminate all FS from the
solution of the three-body Schr\"{o}dinger equation.

\par The three-body Hamiltonian reads:
 \begin{equation}
 \hat{H}=\hat{H}_0
+V(r_{23})+V(r_{31})+V(r_{12}),
\end{equation}
where $\hat{H}_0$ is the kinetic energy operator and $V(r_{ij})$ is
the interaction potential between the  $i$-th and $j$-th particles.
A solution of the Schr\"{o}dinger equation
\begin{equation}
 \hat{H}\Psi_s^{JM}=E\Psi_s^{JM}, \,\, \Psi_s^{JM} \in L_Q.
\end{equation}
should belong to the allowed subspace $L_Q$ of the complete
3$\alpha$ functional space.

The symmetrized wave function of the 3$\alpha$- system is expanded in the series
of Gaussian functions \cite{tur01}:
\begin{equation}
\Psi_s^{JM}=\sum_{\gamma j} c_j^{(\lambda ,l)}\varphi_{\gamma j}^s ,
\end{equation}
where $\varphi_{\gamma j}^s=\varphi_{\gamma
j}(1;2,3)+\varphi_{\gamma j}(2;3,1)+ \varphi_{\gamma j}(3;1,2) ,$
\begin{equation} \varphi_{\gamma
j}(k;l,m)=N_j^{(\lambda l)}x_k^{\lambda}y_k^l\exp(-\alpha_{\lambda j}
x_k^2-\beta_{l j}y_k^2){\cal F}_{\lambda l }^{JM}
(\widehat{\bf{x}_k},\widehat{\bf{y}_k})
\end{equation}
Here $(k;l,m)={(1;2,3),(2;3,1),(3;1,2)}$, $\gamma =(\lambda
,l,J,M)=(\gamma_0,J,M); \bf{x}_k, \bf{y}_k $ are the normalized
Jacobi coordinates in the $k$-set:  $$
\bf{x}_k=\frac{\sqrt{\mu}}{\hbar} (\bf{r}_l-\bf{r}_m)\equiv
\tau^{-1}\bf{r}_{l,m} ; $$ \begin{equation}
\bf{y}_k=\frac{2\sqrt{\mu}}{\sqrt{3}\hbar} (\frac{\bf{r}_l+
\bf{r}_m}{2}-\bf{r}_k)\equiv \tau_1^{-1}\bf{\rho}_k ,
\end{equation}
$N_j^{(\lambda l)}$ is a normalizing multiplier. The nonlinear
variational parameters
 $\alpha_{\lambda j}, \beta_{l j}$ are chosen as the nodes of the
Chebyshev grid:
$$ \alpha_{\lambda j}=\alpha_0\tan(\frac{2j-1}{2N_{\lambda}}\frac{\pi}{2}),
j=1,2,...N_{\lambda}, $$
\begin{equation}
\beta_{l j}=\beta_0\tan(\frac{2j-1}{2N_{l}}\frac{\pi}{2}),
j=1,2,...N_{l},
\end{equation}
where $\alpha_0$ and $\beta_0$ are scale parameters for each
$(\lambda l)$ partial component of the complete wave function.

When one uses the Chebyshev grid, the basis frequensies
$\alpha_{\lambda j}, \beta_{l j}$ cover larger and larger intervals
around the scale parameters as the numbers $N_{\lambda}$ and $N_{l}$
increase. This allows to take into account both short-range and
long-range correlations of particles. The extraordinary flexibility
of the many-particle Gaussian basis makes it possible to describe
three-particle configurations that are formed in the
ground and excited states of multicluster systems, and which
exhibit an extremely high degree of clustering \cite{tur94}.

The angular part of the Gaussian basis is factorized as:
\begin{equation}
{\cal F}_{\lambda l}^{JM}(\widehat{\bf{x}_k},\widehat{\bf{y}_k})=
\{Y_{\lambda}(\widehat{\bf{x}_k}) \bigotimes
Y_l(\widehat{\bf{y}_k})\}_{JM}
 \phi(1) \phi(2) \phi(3),
\end{equation}
where $\phi(i) $ is the internal wave functions of the
$\alpha$-particles. Here the orbital momenta $\lambda$ and $l$ are
conjugate to the Jacobi coordinates $\bf{x}_k$ and $\bf{y}_k$,
respectively.

\par The kinetic energy operator of the Hamiltonian can be expressed
in the normalized Jacobi coordinates in a simple form as
\begin{equation}
\hat{H}_0=-\frac{\partial^2} {\partial\bf{x}_k^2}
-\frac{\partial^2} {\partial\bf{y}_k^2}
\end{equation}
within any choice of $(\bf{x}_k,\bf{y}_k)$, $k=1,2,3$. The detailed matrix
elements of all the operators can be found in \cite{tur94}.

\par In order to separate the complete 3$\alpha$ functional space into
the $L_Q$ and $L_P$ one has to calculate eigen states and corresponding
eigen values of the operator \cite{fuji06}
\begin{equation}
 \hat{P} =\sum_{i=1}^3\hat{P}_i,
\end{equation}
where each $\hat{P}_i, (i=1,2,3)$  is the sum of Pauli projectors
$\hat{\Gamma}_i^{(f)}$ on the partial $f$ wave forbidden states
($1S$, $2S$, and $1D$) in the $i$-th $\alpha \alpha$ subsystem:
\begin{equation}
\hat{P}_i= \sum_{f}\hat{\Gamma}_i^{(f)},
\end{equation}
\begin{equation}
\hat{\Gamma}_i^{(f)}=\frac{1}{2f+1} \sum_{m_f} \mid \varphi_{f
m_f}(\bf{x}_i)> < \varphi_{f m_f}(\bf{x'}_i) \mid
\delta(\bf{y}_i-\bf{y'}_i) ,
\end{equation}
 with the
forbidden state function expanded into the Gaussian basis:
\begin{equation}
 \varphi_{f m_f}({\bf x}_i) =x_i^f \sum_m
N_m^{(f)}b_m^{(f)}\exp(-\frac{r_i^2}{2r_{0m}^{(f)2}})
Y_{fm_f}(\hat{\bf{x}}_i).
\end{equation}
Here $r_0$ is the "projector radius" and $N_m^{(f)}$ is the
normalizing multiplier:
\begin{equation}
 N_m^{(f)}=2^{f+7/4}
\frac{\alpha_m^{(2f+3)/4}}{\pi ^{1/4}[(2\lambda+1)!!]^{1/2}}, \qquad
\alpha_m=\tau^2/(2 r_{0m}^2).
\end{equation}

The operator $\hat{P}$ is not a complete projector
for the three-body system, however its kernel is identical
with the kernel of the complete three-body projector \cite{tur21}.
This is why one can use the operator $\hat{P}$ for the
separation of the complete Hilbert functional space into the allowed
$L_Q$ and forbidden $L_P$ subspaces.

For the study of spatial structure of the $^{12}$C nucleus lowest states  in the 3$\alpha$ model we have to estimate the
probability density functions
\begin{equation}
P(r,R)=\int {\Psi_s^{JM*}({\bf r}, {\bf R} )    \Psi_s^{JM}(\bf{r}, {\bf R} ) } d\widehat{\bf{r}} d\widehat{\bf{R}},
\label{density}
\end{equation}
for each of the fixed bound or resonance state.
In the last equation the integral is taken only over angular part of the relative coordinates $ \bf{r}$ (distance between any fixed two $\alpha$ particles)  and $\bf{R}$ (distance between the last $\alpha$ and the center of mass of already fixed two $\alpha$ particles).  The normalization of the spatial density function requires the condition
 \begin{equation}
\int{P(r,R)}dr dR=1.
\end{equation}

\section{Numerical results}
\subsection{Partial Waves Contributions to the Lowest 0$^+$  and  2$^+$  States}

First we estimate the contributions of different three-body partial waves into the energies of the ground and excited states of the $^{12}$C nucleus. For the 0$_1^+$ (ground) and  0$_2^+$ (Hoyle) states the main contributions come from the three-body
channels $(\lambda,\ell)=(0,0), (2,2), (4,4)$  which contain up to 280 Gaussian functions. Convergence is fast due to the use of symmetrized basis functions. Alternatively, one can use a nonsymmetrized basis functions for practical applications of the three-body wave function. In this case the main contributions come from the three-body channels $(\lambda,\ell)=(0,0), (2,2), (4,4), (6,6,), (8,8)$ requiring up to 1008 Gaussian functions for obtaining convergent results comparable with the results of the symmetrized basis.

\begin{table}
\caption{Contributions of different partial waves in $\%$  into the 0$_1^+$ (ground) and  0$_2^+$ (Hoyle) states} \begin{tabular}{|c|c|c|c|c|c|}\hline
  $(\lambda,\ell)$ &  (0, 0)  & (2, 2)  &    (4, 4)   &   (6, 6) &  (8, 8)    \\ \hline
  $0_1^+$ & 28.777     & 35.722  & 34.885     &  0.606  & 0.010     \\    \hline
  $0_2^+$ & 55.012   & 24.120   & 17.008  & 3.418  & 0.442  \\  \hline
 \end{tabular}
\end{table}

\begin{table}
\caption {Contributions of different partial waves in $\%$ into the 2$_1^+$ and  2$_2^+$ (Hoyle analog) states}
\begin{tabular}{|c|c|c|c|c|c|c|c|c|c|c|c|c|} \hline
  $(\lambda,\ell)$ &  (0, 2)  & (2, 0)  &    (2, 2)   &   (2, 4) &  (4, 2) & (4, 4) & (6, 6) & (6, 4) & (4, 6) & (8, 8) & (8, 6) &  (6, 8) \\ \hline
  $2_1^+$ & 7.186  & 7.199 & 44.240 & 0.736 & 0.724 & 39.313 & 0.586 & 0.005 &0.004 & 0.008 & 0 &0    \\    \hline
  $2_2^+$ & 12.680   & 10.588 & 73.444 & 0.044 & 2.797 & 0.062 & 0.125 & 0.101 &0.031 &0.025 &0.044 &0.058  \\  \hline
 \end{tabular}
\end{table}

In Table 1 we give the contributions of different partial waves into the 0$_1^+$ (ground) and  0$_2^+$ (Hoyle) states.
The ground state energy in the deep phase is -19.90 MeV, which was obtained by including the quantum critical eigen state of the Pauli projector into the functional model space of the 0$^+$ states \cite{tur21a}. A big difference of this number from the experimental energy value of -7.275 MeV means that the 3$\alpha$  system can be in this deep phase with a probability much smaller than 1.
Beyond the critical point the energy of the Hoyle state increases  slightly to $E=-0.458$ MeV, which is, however, lower than the experimental energy value $E_{\rm exp}(0_2^+)$ = 0.380 MeV \cite{tur21a}. Then a small three-body potential is needed for the reproduction of the experimental energy.

As can be seen from the table, for the ground state the largest contributions come from the partial waves
$(\lambda, \ell)=(2,2)$ and $(\lambda, \ell)=(4,4)$. The contribution of the $S$ wave is smaller due to the presence of Pauli forbidden states. In contrast to the ground state, the main contribution to the Hoyle state comes from $S$ waves. This result supports an idea of a possible condensation of the $\alpha$ particles in the Hoyle state \cite{rev}.

For the 2$_1^+$ (bound) and  2$_2^+$ (Hoyle analog) states the most important contributions come from the
three-body channels $(\lambda,\ell)=(0,2), (2,0), (2,2), (2,4), (4,2), (4,4)$  which contain up to 437 symmetrized Gaussian functions. Or, alternatively, nonsymmetrized basis functions can be used within the three-body channels $(\lambda,\ell)=(0,2), (2,0), (2,2), (2,4), (4,2), (4,4), (4,6), (6,4), (6,6), (6,8), (8,6), (8,8)$. In the last case a convergent results were obtained with  1292 Gaussian functions.

In Table 2 we give the contributions of different partial waves to the 2$_1^+$ (bound) and  2$_2^+$ (Hoyle analog) states. The energy of the 2$_1^+$ state in the deep phase is -16.572 MeV, which was obtained by including the quantum critical eigen state of the Pauli projector into the functional model space of the 2$^+$ states \cite{tur21a}. As in the case of the ground state, a strong difference of this number from the experimental energy value of -2.836 MeV means that the 3$\alpha$  system can be in this deep phase with a probability much smaller than 1.  Beyond the critical point the energy of the Hoyle analog state increases  slightly to $E=2.279$ MeV, which is close to the experimental energy value $E_{\rm exp}(2_2^+)$ = 2.596 MeV \cite{tur21a}. The same small additional three-body potential, which was used for fitting Hoyle state energy, can reproduce the experimental energy of its analog state \cite{tur21a}.

From Table 2 one can see that for the  2$_1^+$ state the largest contributions come from the partial waves
$(\lambda, \ell)=(2,2)$  and (4, 4). The contributions of the partial $S$ waves $(\lambda, \ell)=(0,2)$ and (2,0) are comparable and smaller due to the precence of Pauli forbidden states. For the  2$_2^+$ Hoyle analog state the dominant contribution comes from the partial wave $(\lambda, \ell)=(2,2)$.

\subsection{Probability Density Functions of the Lowest 0$^+$  and  2$^+$  States}

Now we go to estimate the probability density functions of the lowest 0$^+$  and  2$^+$  states of the carbon nucleus on the base of Eq. (\ref{density}). These functions yield  a spatial structure or a matter distribution in each state.\newpage
\begin{figure}[htb]
\includegraphics[width=1.0\textwidth]{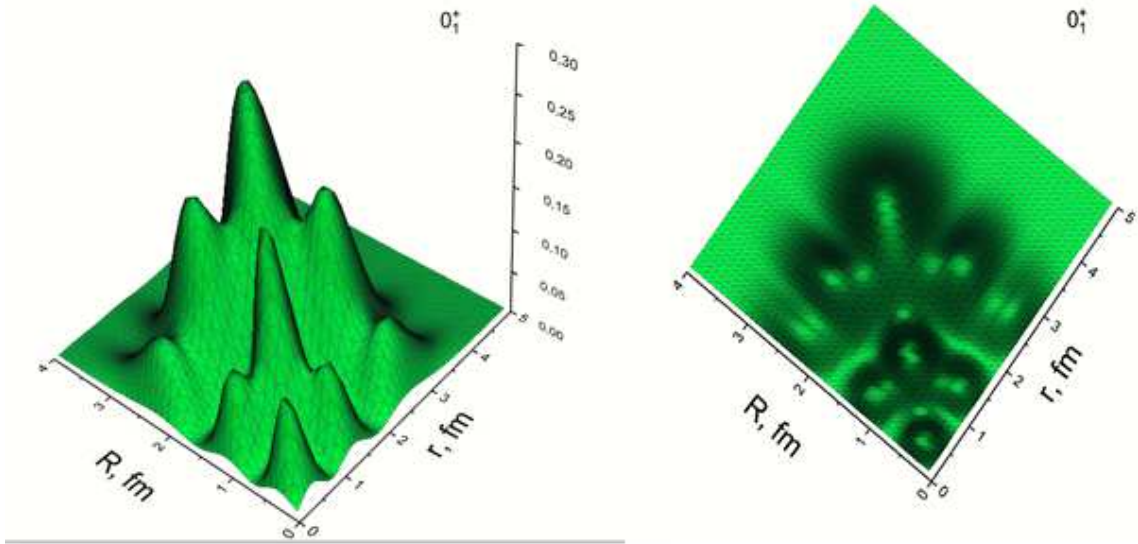}
\caption{Probability density for the $^{12}$C($0_1^+$)  ground state.}
\label{fig_01}
\end{figure}
\begin{figure}[htb]
\includegraphics[width=1.0\textwidth]{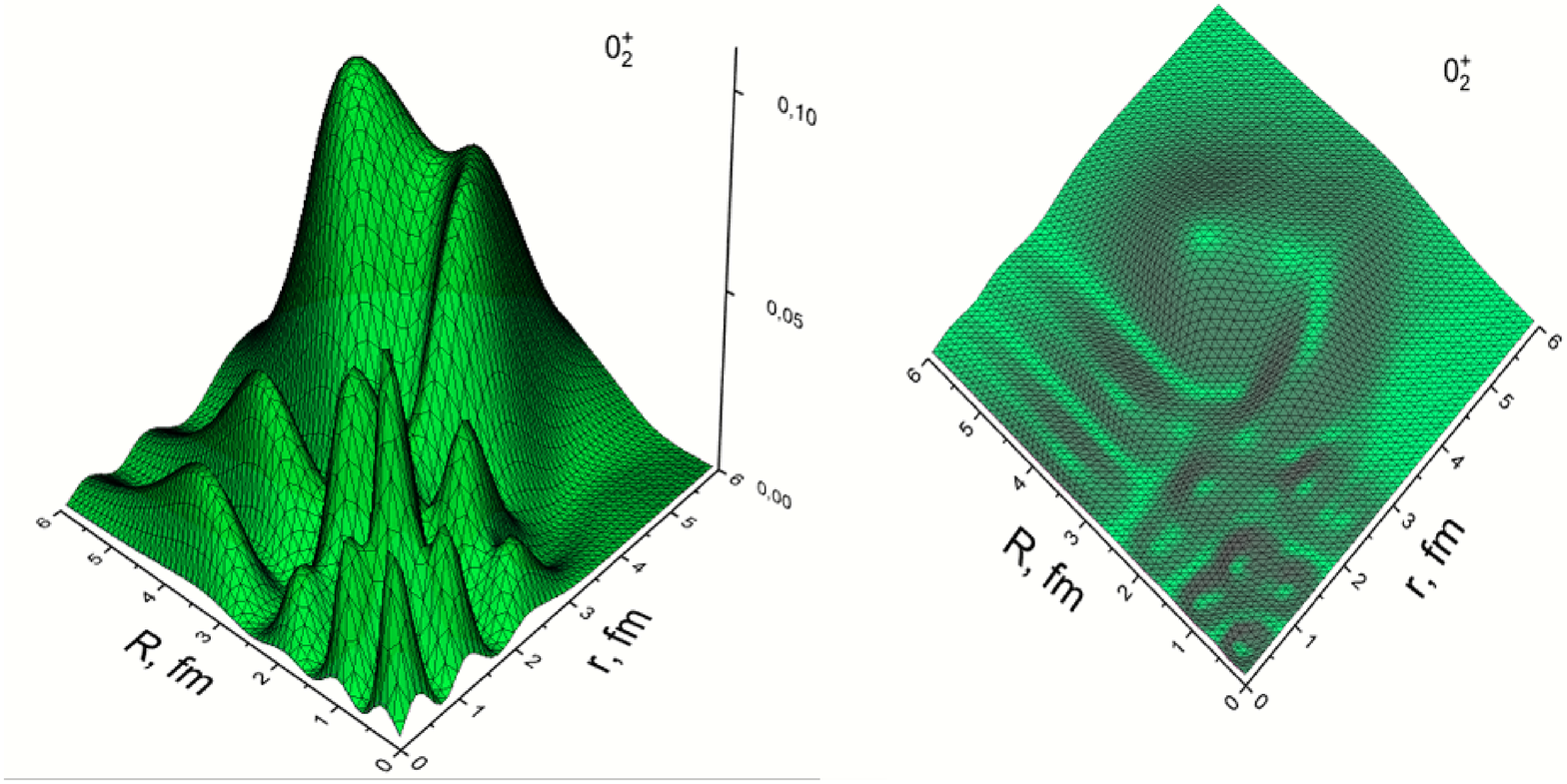}
\caption{Probability density for the $^{12}$C($0_2^+$)  Hoyle resonance state.}
\label{fig_02}
\end{figure}

\begin{figure}[htb]
\includegraphics[width=1.0\textwidth]{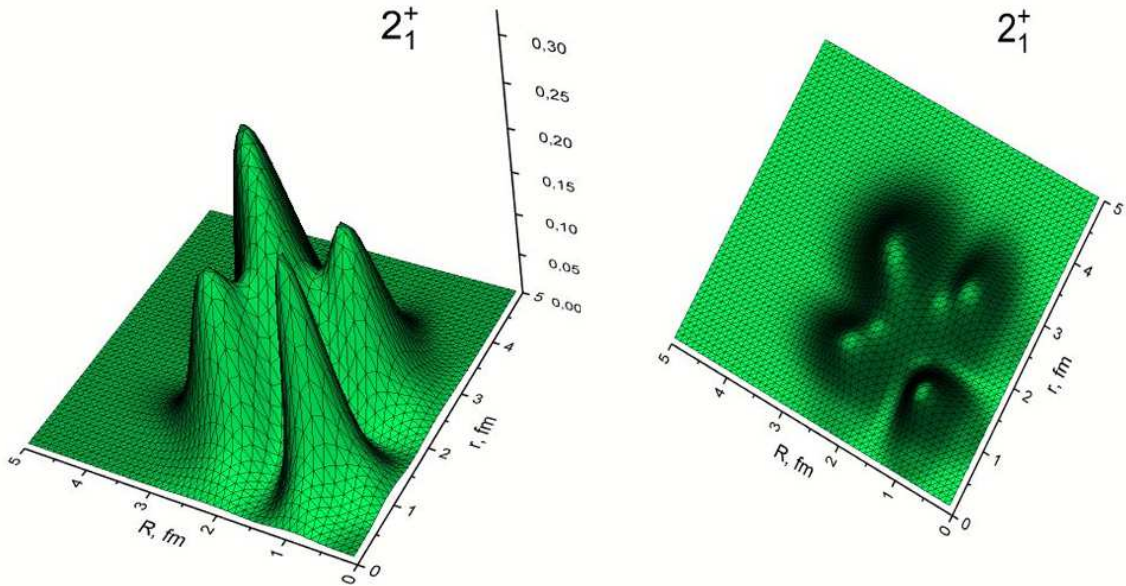}
\caption{Probability density for the $^{12}$C($2_1^+$)  excited bound state.}
\label{fig_21}
\end{figure}
\begin{figure}[htb]
\includegraphics[width=1.0\textwidth]{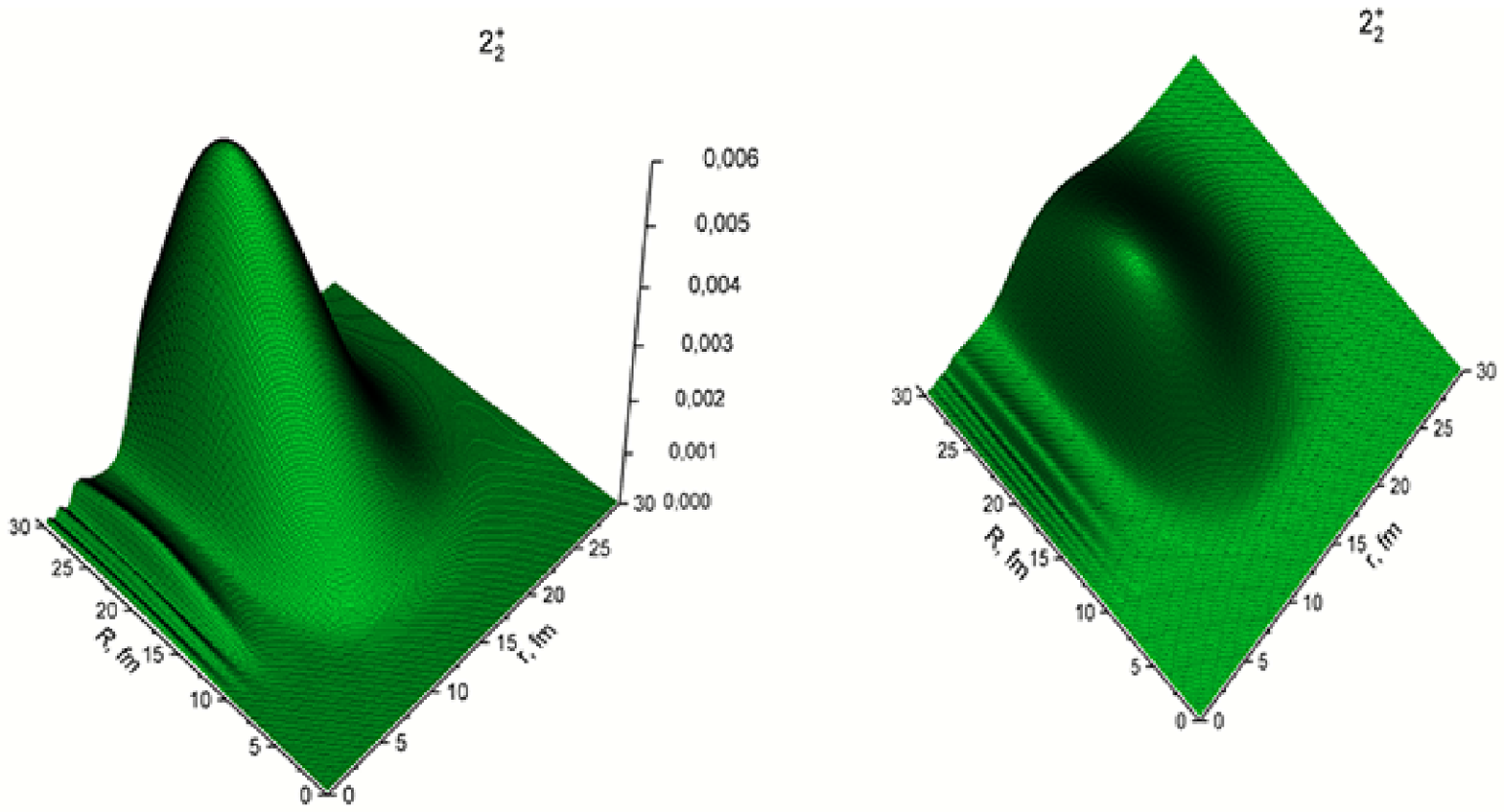}
\caption{Probability density for the $^{12}$C($2_2^+$)  Hoyle resonance analog state.}
\label{fig_22}
\end{figure}
In Fig. \ref{fig_01} we show the probability density functions for the $^{12}$C(0$_1^+$) ground state.
As can be seen from the figure, in this state the alpha particles mostly move in a regular triangular configuration with a maximum probability at the distance of $r=2.7$ fm and $R=2.5$ fm.

In Fig. \ref{fig_02} the probability density function for the $^{12}$C( 0$_2^+$) Hoyle state is presented.
Here one can see a very different structure of the matter distribution. In this state the two alpha particles move at the distance of about $r=$4 fm, and the  last alpha particle moves far from the doublet at the distance between $R=3.0$ fm and $R=5.0$ fm. There is also a probability to go to smaller distances which corresponds to the smaller maxima in the structure function. Here both linear and $^8$Be + $\alpha$ configurations are possible as was found in \cite{sam22}.

In Fig. \ref{fig_21}  the probability density function for the $^{12}$C(2$_1^+$) bound state is demonstrated. As can be seen from the figure, the spatial structure of this state  is  also triangular. It is close to the matter ditribution of the ground state, but  there is a difference at small distances.

In Fig. \ref{fig_22} the probability density function for the $^{12}$C( 2$_2^+$) Hoyle analog state is presented. Here one can see a very extended structure. In this state the two alpha particles move at the distance about $r=$15 fm, and the  last alpha particle moves far from the doublet at the distance between $R=15.0$ fm and $R=30.0$ fm.

\section{Conclusion}

In summary, the spatial structure of the lowest 0$^+$ and 2$^+$ states
of the $^{12}$C nucleus has been
analyzed within the 3$\alpha$ model with the BFW $\alpha \alpha$ potential with two forbidden states in the $S$ wave and a single forbidden state in the $D$ wave. The Pauli forbidden states in the three-body system were treated by the exact orthogonalization method. As was found previously, the ground and excited $2_1^+$ bound states in the deep phases are completely defined by the so-called quantum critical eigen states of the three-body Pauli projector for the 0$^+$ and 2$^+$ functional spaces, respectively. In the weak phases these states are very close to the Hoyle and its analog states.

The contributions of different partial waves into the 0$_1^+$, 0$_2^+$,  2$_1^+$ and  2$_2^+$  states have been estimated.  For the ground  and excited  2$_1^+$  bound states in the deep phases the largest contributions come from the partial waves $(\lambda, \ell)=(2,2)$ and $(\lambda, \ell)=(4,4)$. In contrast to the bound states, for the Hoyle  resonance 0$_2^+$ and its analog state 2$_2^+$, dominant contributions come from the $(\lambda, \ell)=(0,0)$ and $(\lambda, \ell)=(2,2)$  configurations, respectively.

The estimated probability density functions for the $^{12}$C(0$_1^+$) ground and 2$_1^+$ excited bound states show mostly a triangular structure with a distance  of about 2.5 fm between the $\alpha$ particles. However, the spatial structures of the Hoyle resonance and its analog have a strongly different structure, like $^8$Be + $\alpha$.  In the Hoyle state the last $\alpha$ particle moves far from the doublet at the distance between $R=3.0$ fm and $R=5.0$ fm. Even more, in the Hoyle analog state the two alpha particles move at the distance of about 15 fm and the last $\alpha$ particle goes far from the doublet at the distanse between $R=15.0$ fm and $R=30.0$ fm.

\end{document}